\documentclass[referee,useAMS,usenatbib]{mn2e}
\usepackage{graphicx}
\usepackage{rotating}
\usepackage{lscape}

\title[Radio properties of NGC 5903/5898 compact group of galaxies]
  {Radio continuum emission and HI gas accretion in the NGC 5903/5898
compact group of early-type galaxies}
\author[Gopal-Krishna et al.]
  {Gopal-Krishna$^1$\thanks{E-mail:krishna@ncra.tifr.res.in},
   Mukul Mhaskey$^{1}$,
   Paul J. Wiita$^{2}$,
   S. K. Sirothia$^{1}$, 
\newauthor
   N. G. Kantharia$^{1}$ and
   C. H. Ishwara-Chandra$^{1}$
\\
$^1$ National Centre for Radio Astrophysics/TIFR, Pune University Campus, 
Pune 411 007, India\\ 
$^2$ Department of Physics, The College of New Jersey, PO Box 7718, Ewing, 
NJ 08628, USA\\ }
\date{Received 2011 XXXX XXX }

\pagerange{\pageref{firstpage}--\pageref{lastpage}} \pubyear{2011}

\def\LaTeX{L\kern-.36em\raise.3ex\hbox{a}\kern-.15em
    T\kern-.1667em\lower.7ex\hbox{E}\kern-.125emX}

\begin{document}

\label{firstpage}

\maketitle

\begin{abstract}

We discuss the nature of the multi-component radio continuum and HI
emission associated with the nearby galaxy group comprised of two dominant
ellipticals, NGC 5898 and NGC 5903, and a dwarf lenticular ESO514$-$G003.
Striking new details of radio emission are unveiled from the 2nd Data
Release of the ongoing TIFR.GMRT.SKY.SURVEY (TGSS) which provides images with a
resolution of $\sim 24''$x$18''$ and a typical rms noise of 5 mJy at 150 MHz.
Previous radio observations of this compact triplet of galaxies include images
at higher frequencies of the radio continuum as well as HI emission,
the latter showing huge HI trails originating from the vicinity
of NGC 5903 where HI is in a kinematically disturbed state. The TGSS
150 MHz image has revealed a large asymmetric radio halo around NGC 5903
and also established that the dwarf SO galaxy ESO514$-$G003 is the host to a
previously known bright double radio source. The radio emission from NGC 5903
is found to have a very steep radio spectrum $(\alpha \sim -1.5)$ and to
envelope a network of radio continuum filaments bearing a spatial relationship 
to the HI trails. Another noteworthy aspect of this triplet of early-type
galaxies highlighted by the present study is that both its radio loud members, 
namely NGC 5903 and ESO514$-$G003, are also the only galaxies that are seen 
to be connected to an HI filament. This correlation is consistent with the 
premise that cold gas accretion is of prime importance for triggering powerful 
jet activity in the nuclei of early-type galaxies.

\end{abstract}

\section{Introduction}


\vskip0.5in

The E2 type elliptical galaxy NGC 5903 ($z=0.008556$) forms a pair with the E0.5
elliptical galaxy NGC 5898 ($z = 0.007078$); these two are the dominant members of a
compact group that also contains an SO type elongated dwarf 
galaxy ESO 514$-$G003 ($z = 0.007822$; Maia, Da Costa \& Latham 1989; see, also, 
Appleton, Pedlar \& Wikinson 1990, hereinafter APW90). 
The group lies at a distance of $\sim 35$ Mpc (APW90;
Macchetto et al.\ 1996; Rickes, Pastoriza \& Bonatto 2008; Serra \&
Oosterloo 2010), taking a Hubble constant of 70 kms$^{-1}$ Mpc$^{-1}$.
(Komatsu et al.\ 2011). At this distance 1 arcmin corresponds
to a projected distance of 10.2 kpc. The two dominant  galaxies, NGC 5898 
and NGC 5903,
are separated by 5.47 arcmin on the sky, while the third member, ESO514$-$G003
is located 2.95 arcmin south of NGC 5903 (Rickes et al.\ 2008). 
Basic optical and radio parameters of these galaxies are summarized in Table 1. 
In the soft X-ray band, only upper limits are available for their 
luminosities; these limits
are 10$^{40.52}$, 10$^{40.42}$ and 10$^{40.64}$ ergs s$^{-1}$ for NGC 5903, 
NGC 5898 (Beuing et al.\ 1999) and ESO514$-$G003 (Burstein et al.\ 1997), 
respectively.

The first radio observations of this group were reported at
2640 MHz using the twin-element interferometer at the Owens Valley
Radio Observatory, with an east-west spacing of 100 ft, yielding
a flux density of 90$\pm$30 mJy for NGC 5903 (Rogstad \& Ekers 1969).
This was followed up with 5 GHz observations using the Parkes radio
telescope with a pencil-beam of 4 arcmin (Disney \& Wall 1977).
The radio source was found to be well resolved, having a size of $\sim 6$
arcmin and an integrated flux density of 104$\pm$30 mJy, with its peak
closer to the dwarf galaxy ESO514$-$G003. A somewhat clearer picture
emerged when this group was observed in the course of the Ooty lunar
occultation survey at 327 MHz (Gopal-Krishna 1978). The higher resolution
achieved in those observations revealed two resolved sources, the
brighter one coinciding with NGC 5903. The fainter radio source identified
with ESO514$-$G003 was found to have a size of $\sim 1$ arcmin, with its major 
axis inclined eastward from the optical axis of the elongated host SO galaxy.
Also, a comparison of these occultation observations with the available Parkes
pencil-beam measurements (Disney \& Wall 1977) revealed an unusually steep 
spectrum for the
extended radio counterpart of NGC 5903 (Gopal-Krishna 1978; Table 1).
The massive elliptical companion NGC 5898 has, however, remained undetected
in all the radio observations mentioned above, as well as in the later radio
observations (see below).

A major advance in the investigation of this system came from the work
of APW90 who used the VLA at 1.4 GHz to image both the radio continuum (30 arcsec
beam) and HI line emission (45 arcsec beam). Their HI map showed an
HI mass of $\sim 3 \times 10^9 $M$_{\odot}$ to be associated with NGC 5903,
nearly a third of which is in a kinematically disturbed state and distributed
along the galaxy minor axis, while the remainder is in the form of two
huge HI trails extending towards the north and south from the inner HI 
distribution, with projected lengths of approximately 73 kpc and 27 kpc,
respectively. The HI trails are probably the tidal debris left from
an encounter with the disk galaxy ESO 514$-$0050 located 16 arcmin to
the northeast from NGC 5903 (APW90).
Interestingly, while no HI was found associated with the massive
elliptical companion NGC 5898, the southern HI filament was found to
approach the dwarf companion ESO 514$-$0050. The conspicuous lack of both radio
continuum and HI emission from NGC 5898 seemed particularly striking,
when contrasted with the situation for the remaining two members of
the group, namely NGC 5903 and ESO 514$-$G003. Both these galaxies, which
were already known to host extended radio sources (see above), were now
also found to be potential sites of HI gas accretion.

Thus, the apparently correlated presence (or absence) of synchrotron
radio emission and ongoing HI accretion, exhibited by all 3 early-type
galaxies of this group could be a manifestation of the importance of
cold gas accretion for triggering the nuclear jet activity in
elliiptical galaxies, as hypothesized in many papers (Sect. 3).
However, this seemingly self-consistent picture encountered a 
setback due to the assertion of APW90 that the double radio
source (of size $\sim 0.6$ arcmin) which had previously been identified
with the lenticular ESO514$-$G003 in the Ooty lunar occultation survey
mentioned above, was in fact significantly offset from the galaxy and
hence probably a background radio source unrelated to ESO514$-$G003. In
this paper we shall revisit these discrepant findings, using the
recent observations made with the Giant Metrewave Radio Telescope 
\footnote{Operated by the National Centre for Radio Astrophysics (NCRA) of the 
Tata Institute of Fundamental Research (TIFR)} (GMRT, Swarup et al. 1991; 
Ananthakrishnan 2005), which have revealed in unprecedented detail the structure 
of the extended radio synchrotron emission from this compact group of galaxies at 150 MHz. 
In Sect.\ 2 we briefly recapitulate the main optical properties of this 
system and in Sect.\ 3 we present the new radio data.  A discussion and 
conclusions are in Sect.\ 4.

\section{Optical properties of the galaxy triplet}

The salient features of the two dominant members of the group are presented
below.

{\bf NGC 5903:} This galaxy of the morphological type E2 has an
extended red nucleus (Sparks et al.\ 1985).
Macchetto et al.\  (1996) have detected in its inner parts, a
$4 \times 2.5$ kpc$^2$ wide region of ionised gas, extended along the
galaxy major axis. This region encompasses a dusty patch of size
$\sim$3.3 kpc inferred from the $(V - R)$ colour index map, which is
slightly elongated at P.A. $\sim 135^{\circ}$ (Ferrari et al.\ 1999).
According to Rickes et al.\ (2008), the observed MgII abundance gradient is
consistent with the origin of this massive galaxy, of dynamical mass
$\sim 2 \times 10^{11}$ M$_\odot$, in merger events. Their 
stellar population synthesis has revealed at least two stellar populations
with ages of 13 Gyr and 5 Gyr and, possibly, a younger population of age
$\sim 1$ Gyr, which is mainly discernible in the central parts of this
elliptical. 
Deep V-band imaging has revealed a modest degree of tidal disturbance
in the stellar body of this galaxy, which is typical of nearby massive
ellipticals in groups, which are thought to grow mainly via ``dry"
mergers (Tal et al.\ 2009). However, Serra \& Oosterloo (2010) have
argued that the presence of cold gas may often play a significant
role in the assembly of massive galaxies outside cluster environments.
Finally, slit spectra taken along the optical major axis of this
elliptical show little rotation of the stellar body, while the ionized
gas has an irregular velocity profile along the radius (Caon, Macchetto
\& Pastiroza 2000; also Longo et al.\ 1994).

{\bf NGC 5898:} This nearly circular E0.5 elliptical shows an outer arc
of dust and an inner red patch of elliptical shape. The gap between
them indicates either intermittent accretion or a rapid dissipational
process (Sparks et al.\ 1985). Macchetto et al.\ (1996) have detected
a roughly circular central patch of ionised gas of diameter $\sim 6$ kpc.
A deep V-band image shows tidal disturbance that is much stronger
than is typical of nearby massive ellipticals in groups (Tal et al.\ 2009),
and thus suggestive of a major merger event. This is consistent with the complex
kinematics revealed by the detection of a counter-rotating stellar
core on the galaxy major axis (PA $= 62^{\circ}$), with a transition
radius of 1 kpc and a gas rotation which, remarkably, is opposite to
that of the stars on the galaxy minor axis (Caon et al.\ 2000; Bertola
\& Bettoni 1988).

\section{Radio emission from the galaxy triplet system}
Figure 1 reproduces a new radio map of this compact group, taken from
the 2nd Data Release (DR2) of the ongoing 
TIFR.GMRT.Sky.Survey\footnote{http://tgss.ncra.tifr.res.in/} (TGSS)
which is in the process of imaging the entire sky north of declination
$-53^{\circ}$  at 150 MHz with a resolution of $\sim 20$ arcsec and a 
typical rms noise of 5--7 mJy/beam (the value for the present map
is on the lower side, being only 3 -- 4 mJy/beam).
The radio contours are overlaid on the R-band optical photograph 
reproduced from the Digital Sky 
Survey\footnote{http://archive.stsci.edu/dss/}. In Figure 2 we 
show the same radio image in grey scale, overlaid on which are the 
outermost (lowest) 4 contours of HI emission, reproduced from the VLA 
map made by APW90. Lastly, we show in Figure 3 the 150 MHz (TGSS) and 
1.4 GHz (NVSS) emission contours in the region of the dwarf SO 
galaxy ESO514$-$G003.

The TGSS map at 150 MHz has unravelled, for the first time, the complexity
of the radio emission associated with this group of 3 early-type galaxies,
although the massive elliptical NGC 5898 is still undetected.
The 150 MHz image of the other massive elliptical, NGC 5903, exhibits a number 
of interesting features, including a jet-like structure of size $\sim$ 1 
arcmin and flux density 350 $\pm$ 65 mJy, protruding from the elliptical 
towards the southwest. The galaxy is surrounded by a diffuse radio halo of 
diameter $\approx$ 7.5 arcmin (77 kpc) and flux density $\approx$ 6.7 
$\pm$ 1.1 Jy encompassing a web of radio filaments and extending over 
the region bounded by the galaxy triplet. Since the filaments
have not been discerned in the previously published radio maps at higher 
frequencies, probably due to their lower resolution (Sect.\ 1), it is
presently not possible to determine their radio spectrum. 
Nonetheless, the available data do provide reliable estimate of the
integrated radio spectrum of this elliptical NGC 5903 between 150 MHz
and 1.4 GHz where the maps have adequate resolution (Table 1). Although 
the highest frequency at which this source has been mapped is 5 GHz, 
the published value of 104 $\pm$ 30 mJy measured with the 4 arcmin 
Parkes pencil-beam can only be regarded as a generous upper limit, since 
much of this emission is contributed by the dwarf neighbour ESO 514$-$G003, 
as inferred from the marked shift in the radio centroid at 5 GHZ towards 
it (see Table 2 of Disney \& Wall 1977). 
The spectral index for NGC 5903 is ultra-steep ($\alpha = -1.5\pm0.08$,
with $S_{\nu} \propto \nu^{\alpha}$), 
in agreement with the previous estimate (Gopal-Krishna 1978).

The 150 MHz map is also useful for clarifying the confusion over the 
radio emission from ESO 514$-$G003, the least massive, lenticular member 
of this triplet.  As mentioned in Sect. 1, APW90 found the bright symmetric
double radio source in their 1.4 GHz VLA map (resolution $30^{\prime \prime}$) to be 
significantly displaced from the optical galaxy; they concluded 
that it is an unrelated background radio source (with an undetected optical
counterpart). On the other hand, the lunar occultation scans at 327 MHz
were consistent with a physical association of the radio source with
the lenticular galaxy (Gopal-Krishna 1978). To pursue the matter, we
have overlaid on the galaxy image the contours of radio emission at 
150 MHz (TGSS) and 1.4 GHz (NVSS), also marking the position of the 
mid-point of the symmetric double radio source seen in the 1.4 GHz
VLA map by APW90 (their Fig. 7b). As seen from Figure 3, the position 
offset of the galaxy ESO514$-$G003 from the TGSS position of the centre 
of the double radio source at 150 MHz is 8 arcsec (which is at least a 
factor of two smaller than the offset of the galaxy from the APW90 position 
mentioned above), which is consistent with the expected TGSS position uncertainty,
given the confusion arising from the radio halo of NGC 5903. Likewise, a 
reasonable agreement of the galaxy position is also found with the NVSS 
position at 1.4 GHz, though the radio source is poorly resolved due to 
the larger beam. Thus, based on both TGSS (150 MHz) and NVSS (1.4 GHz) 
positions we conclude that the double radio source is associated with 
the lenticular ESO514$-$G003, in agreement with the lunar occultation 
results at 327 MHz (Sect.\ 1; Gopal-Krishna 1978). Note the westward extension 
of the double radio source visible in the TGSS map raises the possibility 
that the radio source may actually be a triple source (with a flux 
density of $\sim$ 1150 $\pm$ 190 mJy at 150 MHz), in which case the agreement with 
the galaxy position would be even better.

\section{Discussion}

Based on their 1.4 GHz VLA observations, APW90 characterised the radio
counterpart of NGC 5903 as a head-tail radio source, with the tail
extending to the southwest. Such a radio morphology seems puzzling,
since the host galaxy belongs to a very small group sparse in hot
intra-group gas (Sect. 1). The present TGSS map at 150 MHz has revealed
faint radio emission on the opposite, northeastern side of the galaxy
as well. The radio source may thus be an
asymmetric radio halo around NGC 5903, with most of the radio emission 
located on the south-western side of the galaxy, within the region 
demarcated by the galaxy triplet (Figure 1). The combination of high 
sensitivity and resolution of the 150 MHz TGSS map has revealed an
extensive network of radio filaments, the brightest of which fills
the space delineated by the prominent HI trails which are seen to
extend towards the south and southwest from NGC 5903, in a `bay' like
configuration (Figure 2).   As the filamentary structure has not been 
resolved at any other frequency, its radio spectrum is presently unknown.  
We shall make a plausible assumption that the radio emission is 
synchrotron radiation.   Figure 2 shows the morphological relationship of 
the radio continuum filaments to the HI trails reproduced from APW90.
Based on the observed HI velocity gradients and other arguments, APW 
have estimated that the huge HI filaments are transient features that
would have dissipated had they been older than $\sim 3 \times 10^8$ 
yrs. The ultra-steep spectrum of the radio source ($\alpha \sim - 1.5$, 
Table 1) suggests that the source is very old and its age is likely to 
be of order of the maximum lifetime of the radio mode (roughly $10^8$ yr, 
e.g., Leahy, Muxlow \& Stephens 1989; Liu et al.\ 1992; Barai \& Wiita 2006; 
Croton et al.\ 2006; Shabala et al.\  2008). 
Thus, the triggering of the currently observed radio
source in NGC 5903 may well have coincided with the formation of the HI 
trails and cold gas accretion through them, onto the galactic nucleus.
This situation is rare and contrasts with the usual pattern found for 
nearby radio-loud elipticals, wherein the detected radio source is much 
younger than the associated extended HI structure (typically a regular 
rotating disk, or ring-like formation, stretching beyond the optical
extent of the host elliptical) (e.g., Tadhunter et al.\ 2005; 
Emonts et al.\ 2006). This has led to the view that if any 
radio activity was triggered at the time of a (major) merger between two 
gas-rich galaxies, resulting in an elliptical (e.g., Naab, Burkert \& 
Hernquist 1999), the relic of that radio source is generally not visible
any more (e.g., Emonts et al.\ 2007). Note that the latest
major starburst episode in this elliptical is estimated to have occurred
approximately 1 Gyr ago (Rickes et al.\ 2008), which  
presumably predates the birth of the radio source we see today. However, based on the
recent evidence, the link between starburst and radio-loud nuclear 
activity does not seem to be strong (Dicken et al.\ 2011).
 
Further, this nearby galaxy provides a striking manifestation of the 
influence of the HI tidal trails on the radio morphology. From 
Figure 2 the brightest radio filament is aligned with the inner (eastern) edge 
of the southwestern bright HI trail. This alignment, together with the 
lateral offset from the HI ridge-line, appears to signify a 
dynamical interaction of the HI trail (presumably infalling onto NGC 5903)
with the surrounding synchrotron plasma that led to compression of the 
magnetic field along its eastern edge.
This process could boost the radio emission locally, if relativistic particles 
are entangled in the regions of compressed magnetic field associated 
with the HI filaments. 
Such a `trapping' of the synchrotron plasma by the HI filaments amounts
to a kind of prolonged `inertial containment' of the synchrotron
plasma, which could explain the ultra-steep spectrum of the diffuse 
radio source at metre wavelengths, even though such spectrum is generally not expected 
for sources outside clusters of galaxies, where the confining hot 
gaseous environment is lacking (Sect.\ 1).

Lastly, a highly suggestive correlation emerging from the present 
work derives from our establishing the radio counterpart to the 
SO galaxy ESO514$-$G003. With this, two members of this triplet of
early-type galaxies are shown to have radio counterparts and these 
two galaxies are also the only ones that appear connected to HI trails 
and therefore likely to be accreting the cool HI gas. The third member, 
the dominant elliptical NGC 5898 is radio quiet and also unconnected to 
any HI trail (APW90). With the obvious caveat of small number statistics
(such a relation has a 1/8 chance of occurring randomly in the present case,
if both HI gas and radio emission are independent and each have a probability of 
$0.5$), this correlation between radio loudness of an early-type galaxy 
and its physical 
connection to an HI filament (or, vice versa), seems entirely consistent 
with the importance of cold gas accretion in triggering and sustaining 
powerful jet activity in the galactic nuclei, as highlighted in the 
recent literature (e.g., Best et al.\ 2005; Croton et al.\ 2006). 
It is tempting to speculate that the correlation with radio emission 
arises because, in comparison to ionized gas (viz., the case of Bondi 
accretion: Allen et al.\ 2006; Hardcastle et al.\ 2007; Balmaverde 
et al.\ 2008; Buttiglione et al.\ 2010), cold gas is more 
amenable to fast accretion merely because any infalling cold gas 
would encounter much less resistance from the substantial magnetic 
fields likely to exist within the cores of massive galaxies.
We are conducting a search of other small groups mapped by TGSS that
also have HI data available, to investigate whether the correlation
between HI filaments and radio activity in early-type galaxies seen in
this particular triplet remains valid.

\section*{Acknowledgments}
The Giant Metrewave Radio Telescope (GMRT) is a national facility operated by the National Centre for 
Radio Astrophysics (NCRA) of the Tata Institute of Fundamental Research (TIFR). 
We thank the staff at NCRA and GMRT for their support. This research has used the
TIFR. GMRT. Sky. Survey (http://tgss.ncra.tifr.res.in) data products,
NASA's Astrophysics Data System and NASA/IPAC 
Extragalactic Database (NED),
which is operated by the Jet Propulsion Laboratory, California Institute
of Technology, under contract with National Aeronautics and Space Administration.


\newpage
\begin{landscape}
\begin{table}
\small
\centering
\caption{: Basic parameters of the galaxy triplet.$^{\ast}$}
\begin{tabular}{cccccccccccc}\\
\hline
\multicolumn{1}{c}{Galaxy} & \multicolumn{1}{c} {Type} &\multicolumn{1}{c}{m$_{V}$} & \multicolumn{2}{c}{Galaxy Coordinates{$\ddagger$}} & \multicolumn{4}{c} {Flux desnities and errors (rms)} & \multicolumn{1}{c}{$\alpha_{150}^{1400}$ } & \multicolumn{1}{c}{Luminosity}  \\
 & & & $\alpha$ & $\delta$ & 150 MHz & 327 MHz & 1.4 GHz & 5 GHz &  & at 150 MHz \\
 & & & (J2000) & (J2000) & (TGSS)& (Ooty) & (NVSS) & (Parkes) &  &  \\ 
 & & &h   m   s &   $^{\circ}$ $^{\prime}$ $^{\prime\prime}$          &  mJy & mJy & mJy & mJy & & W.Hz$^{-1}$ \\
\hline
NGC 5903 & E2 & 11.74 & 15 18 36.1 & -24 04 07 & 7067${\pm}$1100 &1300${\pm}$250& 261${\pm}$15& $104^{\dagger}$${\pm}$30& -1.5${\pm}$0.08 & 1.03x$10^{24}$\\
ESO514$-$G003 & S0 & 13.83 & 15 18 35.1 & -24 07 12 & 1013${\pm}$160& 700${\pm}$200& 105${\pm}$4& &-1.0${\pm}$0.07 & 1.47x$10^{23}$ \\
NGC 5898 & E0.5 & 11.76 & 15 18 13.6 & -24 05 52 & $-$ & $-$ & $-$ & $-$ & $-$& $-$\\
\hline
\end{tabular}

{$\ast$} Distance = 35 Mpc (Sect. 1) \\
{$\ddagger$} Taken from NED \\
{$\dagger$} Also includes flux density of ESO514$-$G003 \\

\end{table}
\end{landscape}

\clearpage

\begin{figure*}
\hspace*{-1.0cm}
\hbox{
\includegraphics[height=15.0cm,width=15.0cm]{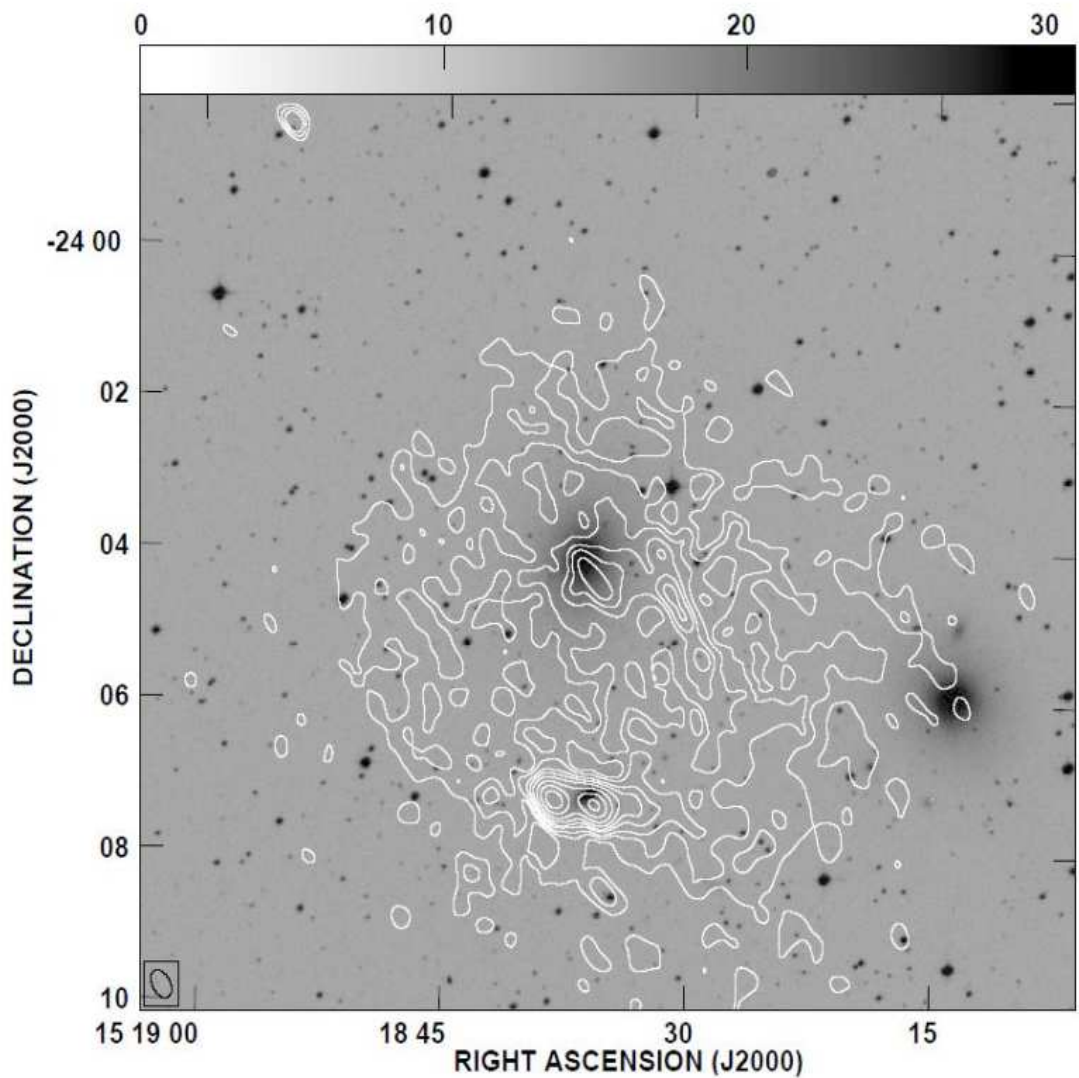}
}
\caption{Contours of 150 MHz (TGSS) radio continuum emission overlaid on 
the optical DSS image (R-band) of the galaxy triplet, showing NGC5898 to 
the west, NGC5903 to the north and ESO514$-$G003 to the south. The FWHM of 
the synthesized radio beam is plotted in the lower left corner (see text).
The contour levels are: 2,3,4,6,8,11,16,23,31,44 (in units of 7.17 mJy/beam)} 
\label{fig:1}
\end{figure*}
\clearpage

\begin{figure*}
\hspace*{-1.0cm}
\hbox{
\includegraphics[height=18.0cm,width=15.0cm]{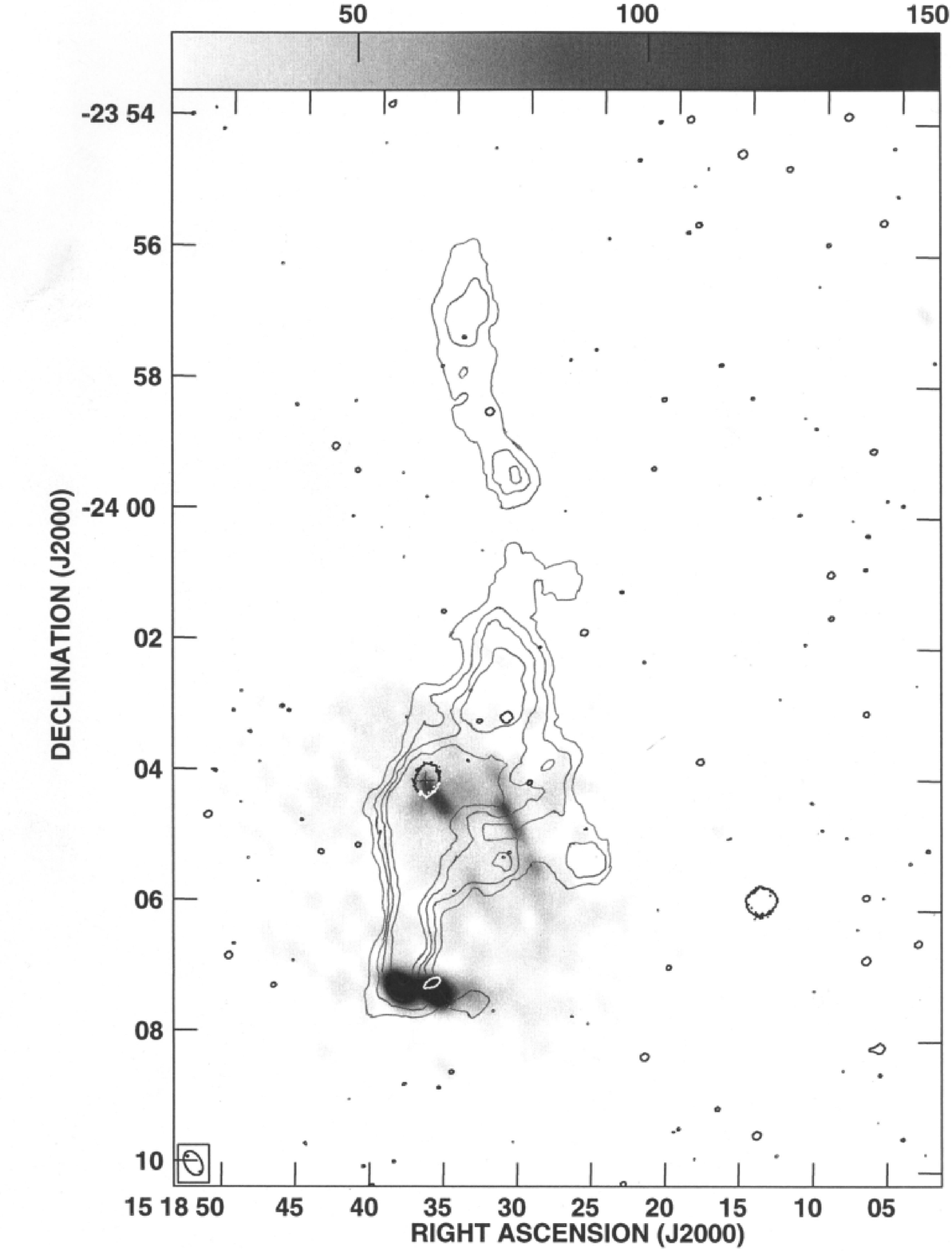}
}
\caption{The lowest 4 contours of HI surface density 
(from Appleton et al.\ 1990) overlaid on the 150 MHz continuum image (grey scale). 
The profiles of the three early-type galaxies of this triplet are shown as ellipses, 
as described in Figure 1. Contour units are 0.6,1.2,1.7 and 2.3 x$10^{20}$ atom cm$^{-2}$}  
\label{fig:2}
\end{figure*}
\clearpage

\begin{figure*}
\hspace*{-1.0cm}
\hbox{
\includegraphics[height=18.0cm,width=15.0cm]{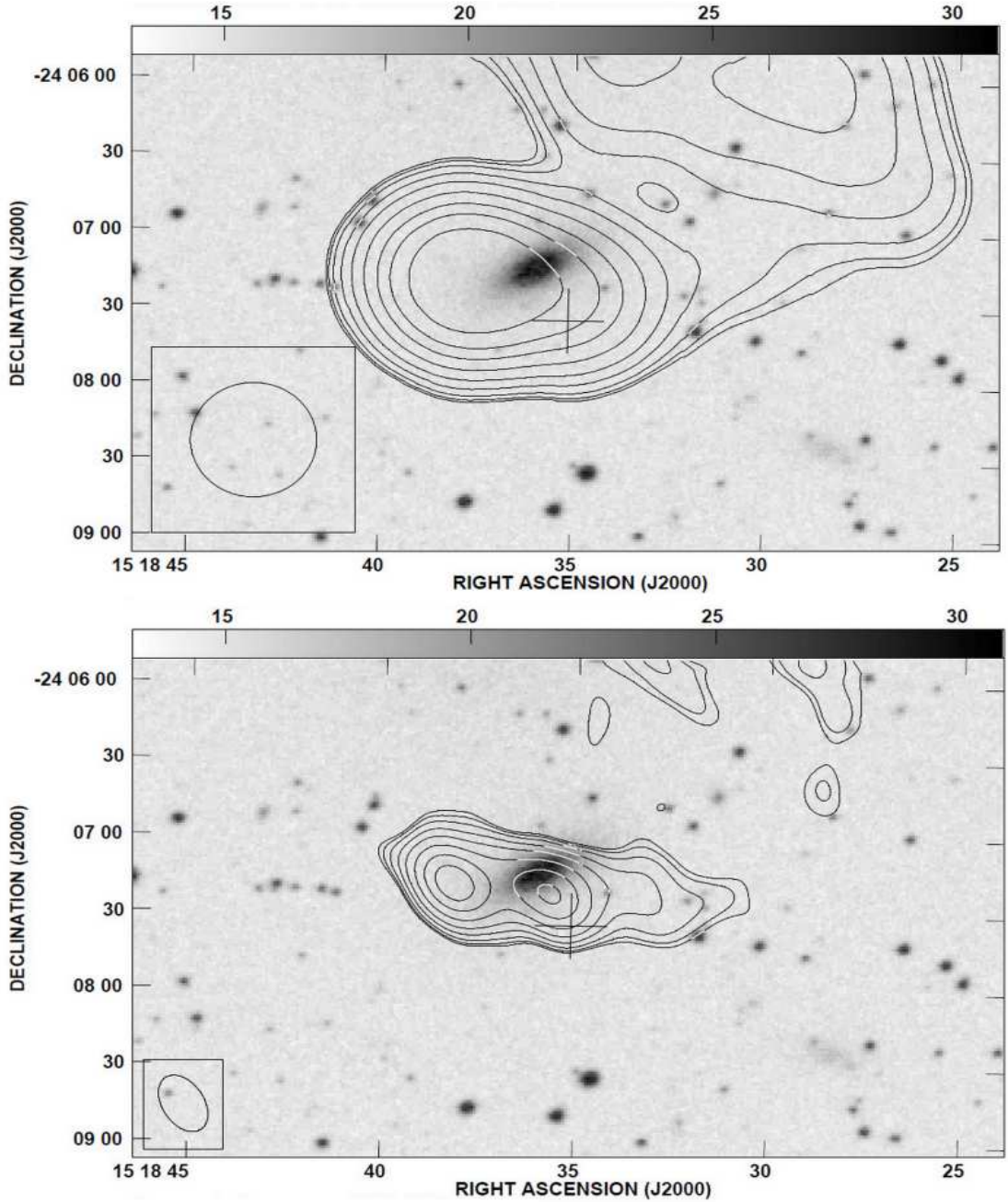}
}
\caption{Contours of radio continuum emission are shown overlaid on the optical
DSS (R-band) image of the SO galaxy ESO514$-$G003. The upper panel shows the
1.4 GHz emission (NVSS) while the lower panel shows the emission contours
at 150 MHz (TGSS). The synthesized beam is shown in the lower left
corner, of each panel. The cross in each panel marks the centre of the symmetric 
double radio source seen in Fig. 7(b) of Appleton et al (1990).
The contour levels are: 2,3,4,6,8,11,16,23,31,44 (in units of 7.17 mJy/beam)}
                
\label{fig:3}
\end{figure*}
\clearpage

\end{document}